\documentclass[preprint]{revtex4}

\usepackage{graphicx}

\begin{document}

\title{Pulsar Magnetosphere: Variation Priciple, Singularities, Estimate of Power}

\author{Andrei Gruzinov}
 
\affiliation{Center for Cosmology and Particle Physics, Department of Physics, New York University, NY 10003}

\date{October 25, 2005}

\begin{abstract}

We formulate variation principle for force-free magnetosphere of an inclined pulsar: ${\cal E} +{\bf \Omega}\cdot {\bf M}$ ( $\cal E$, ${\bf M}$ are electromagnetic energy and angular momentum, ${\bf \Omega}$ is the angular velocity of a star) is stationary under isotopological variations of magnetic field and arbitrary variations of electric field. The variation principle gives the reason for existence and proves local stability of current singular layers along magnetic separatrices. Magnetic field lines of inclined pulsar magnetosphere lie on magnetic surfaces, and do have magnetic separatrices. 

In the framework of the isotopological variation principle, inclined magnetospheres are expected to be simple deformations of the axisymmetric pulsar magnetosphere. A singular line should exist on the light cylinder, where inner separatrix terminates and outer separatrix emanates. The electromagnetic field should have  an inverse square root singularity near the singular line inside the inner magnetic separatrix. 

Large distance asymptotic solution is calculated, and used to estimate the pulsar power, $L\approx c^{-3}\mu ^2\Omega^4$ for spin-dipole inclinations $\lesssim 30^{\circ}$ . 

\end{abstract}

\maketitle

\section{Introduction}

Pulsars are very simple and clean astrophysical objects. For all we know, pulsars are just magnetized spinning conductors. The only parameters of a pulsar are the distribution of entering magnetic field over the surface of a star and the angular velocity. For some observables, the magnetic dipole might be the only relevant input parameter characterizing the magnetic field. For these observables, pulsars form just a three-parameter family of objects -- all pulsars with the same spin, dipole, and spin-dipole angle are equal. Yet the beautifully rich pulsar phenomenology remains unexplained quantitatively: given spin and dipole, we cannot predict radiation.

For young pulsars, like Crab, the spin-down power is much greater then the power of observed electromagnetic radiation (eg. Fig. 9 in \cite{kuiper}). This might indicate that the magnetosphere is close to a force-free state, and the magnetosphere can be calculated independently of radiation. Once the force-free magnetosphere is understood, one will be on firmer ground for predicting radiation (calculating small dissipative corrections to a pure force-free state). In any case, quantitative predictions of radiation require the knowledge of magnetosphere.

The quantitative theory of force-free pulsar magnetosphere has recently appeared. Based on classical early insights \cite{goldreich, scharlemant, michel}, the shape, singularity structure, spin-down power, and MHD-realizability of an axisymmetric pulsar (meaning pulsar with small spin-dipole inclinations) has been established \cite{contopoulos, gruzinov, komissarov}. In particular, the spin-down power of an axisymmetric pulsar turned out to be 1.5 times greater than perpendicular dipole power:  $L\approx c^{-3}\mu ^2\Omega^4$. 

Force-free magnetosphere of an inclined pulsar should be a doable problem too. Here we show that inclined pulsar magnetosphere should possess stable current/charge layer singularities. This is done using isotopological variation principle (\S2 ). We also calculate the large-distance asymptotic of the magnetosphere, and use this asymptotic solution to give a crude estimate of pulsar power (\S3 ). In Appendix, we discuss a mathematical problem of minimizing functionals by isotopological transformations. We show that isotopological relaxation turns separatrices into singular current layers.

\section{Isotopological variation principle and singularities}

We assume that magnetosphere is force free: 
\begin{equation}
\rho {\bf E}+{\bf j}\times {\bf B}=0.
\end{equation}
Using Maxwell equations
\begin{equation}
\partial _t{\bf B} =-\nabla \times {\bf E},~~~ \rho =\nabla \cdot {\bf E},~~~{\bf j}=\nabla \times {\bf B}-\partial _t{\bf E},
\end{equation}
assuming that electromagnetic pattern rotates with velocity ${\bf V}={\bf \Omega}\times {\bf r}$, just like it does for a dipole in vacuum:
\begin{equation}
\partial _t{\bf B} =\nabla \times ({\bf V}\times {\bf B}),~~~\partial _t{\bf E} ={\bf E}\cdot \nabla {\bf V}-{\bf V}\cdot \nabla {\bf E},
\end{equation}
and using the condition of zero tangential electric field on the surface of the star in the star frame, we get ${\bf E}=-{\bf V}\times {\bf B}$ and the pulsar magnetosphere equation
\begin{equation}\label{basic}
{\bf B}\times \nabla \times \left( {\bf B} + {\bf V}\times ({\bf V}\times{\bf B})\right) =0.
\end{equation}
 
The pulsar magnetosphere equation (\ref{basic}) follows from a variation principle: $\delta S=0$, where $S$ is proportional to electromagnetic action
\begin{equation}\label{vary}
S=\int d^3r (B^2-({\bf V}\times{\bf B})^2),
\end{equation}
and the variation of action should be calculated for isotopological perturbation of magnetic field 
\begin{equation}\label{isotop}
\delta {\bf B}=\nabla \times (\delta \vec{\xi }\times{\bf B}).
\end{equation}
In other words, action should be extremized by continuously displacing magnetic field lines. 

The isotopological variation requirement is more than just a mathematical device giving stationary solutions. The actual time evolution of magnetic field in force-free electrodynamics is isotopological, because $\partial _t{\bf B} =\nabla \times ({\bf v}\times {\bf B})$ (\cite{gruzinov}, ${\bf v}\equiv {\bf E}\times {\bf B}/B^2$). Then locally, the variation principle requires minimizing proper magnetic energy (magnetic energy in a frame where the electric field is absent) by displacing the magnetic field lines. This immediately shows that current layers are locally stable. 

An equivalent formulation of the variation principle (\ref{vary}, \ref{isotop}) states that for stationary magnetosphere, the first variation of the following integral of motion of FFE (force-free electrodynamics, Appendix in \cite{gruzinov})
\begin{equation}
{\cal E} +{\bf \Omega}\cdot {\bf M}= \int d^3r (~{B^2+E^2\over 2}+{\bf \Omega}\cdot {\bf r}\times ({\bf E}\times{\bf B})~)
\end{equation}
vanishes. Here the electric field variation $\delta {\bf E}$ is arbitrary and the magnetic field variation $\delta {\bf B}$ is itopological, given by (\ref{isotop}).

In the framework of isotopological variation principle, the only difference between inclined and axisymmetric pulsar magnetospheres is the distribution of entering magnetic field lines over the surface of the star. It is then plausible that stationary inclined magnetospheres do exist.

Another way to see that stationary inclined magnetospheres may exist, is to write the magnetosphere equation (\ref{basic}) as
\begin{equation}\label{basic2}
\nabla \times \left( {\bf B} + {\bf V}\times ({\bf V}\times{\bf B})\right) =\lambda {\bf B},
\end{equation}
where $\lambda$ is constant along field lines: ${\bf B}\cdot \nabla \lambda =0$. One needs to specify the scalar $\lambda$ on the surface of the star. Just like for an axisymmetric dipole \cite{contopoulos}, the right choice of $\lambda$ on the surface of the star may  allow to smoothly cross the light cylinder (light cylinder is the surface $V=1$).

Magnetosphere equation (\ref{basic2}) shows that stationary magnetospheres possess magnetic surfaces. This is because ${\bf B}\cdot \nabla \lambda =0$, meaning that magnetic field lines lie on the surfaces $\lambda =$const. Magnetic separatrix surfaces will then exist, between open and closed field lines within the light cylinder, and between oppositely directed field lines outside the light cylinder. As we show in  Appendix, isotopological extrema are expected to have stable singular current layers along magnetic separatrices. 

Moreover, in view of the variation principle, inclined magnetospheres should be just simple deformations of the axisymmetric magnetosphere, allowing to propose the following structure of the pulsar magnetosphere, based on the singularity structure found in the axisymmetric case: 

\begin{enumerate}
\item There is a singular line on the light cylinder.
\item Inner separatrix terminates on the singular line.
\item Outer magnetic separatrix emanates from the singular line.
\item Separatrices are singular current/charge layers.
\item The inner separatrix has a wedge at the singular line.
\item In the lab frame, the electromagnetic field has an inverse square root singularity near the singular line inside the inner separatrix.
\end{enumerate}

This picture should be checked by an accurate numerical simulation of the inclined magnetosphere. In the next section, we will just show that an asymptotic solution of the magnetosphere equation exists. We use this asymptotic solution to give a rough estimate of pulsar power.

\section{Pulsar Asymptotic and Power}

At large distances from the light cylinder (for $V\gg 1$), the magnetosphere equation (\ref{basic2}), written in spherical coordinates $(r,\theta ,\phi)$, with $\vec{ \Omega}$ along $\theta =0$, in units $\Omega =c=1$, is

\begin{equation}
\partial _{\theta}(\sin \theta B_{\phi})+r^2\sin ^2\theta \partial _\phi B_\theta =r\sin \theta \lambda B_r,
\end{equation}
\begin{equation}
\partial _r(rB_\phi )+r^2\sin \theta \partial _\phi B_r=-r\lambda B_\theta ,
\end{equation}
\begin{equation}
\sin ^2\theta \partial _r(r^3B_\theta )-r^2\partial _\theta (\sin ^2\theta B_r)=-r\lambda B_\phi ,
\end{equation}
\begin{equation}
\sin \theta \partial _r(r^2B_r)+r\partial _\theta(\sin \theta B_\theta )+r\partial _\phi B_\phi =0.
\end{equation}

These equations have the following asymptotic solution:
\begin{equation}
B_r=fr^{-2}+o(r^{-2}),~~B_\theta =gr^{-3}+o(r^{-3}), ~~B_\phi =-\sin \theta fr^{-1}+o(r^{-1}),
\end{equation}
where $f$, $g$ are functiuons of $\theta$ and $\tilde{\phi }$, with $\tilde{\phi }\equiv \phi +r$. As it should be, asymptotically one has $E=B$, and Poynting flux $S\propto r^{-2}$. 

One can use the asymptotic solution for a crude, but ``physically motivated'' estimate of the pulsar power. Using the asymptotic solution, we calculate the pulsar power (Poynting flux through a distant sphere) as
\begin{equation}\label{power}
L\propto \int d\Omega f^2\sin ^2\theta \propto \int dN (dN/d\Omega )\sin ^2\theta ,
\end{equation}
where $\Omega$ is the solid angle, and $N$ is the number of field lines. While we obviously cannot calculate this integral without a full solution, we can offer the following speculation. 

The number of field lines going to infinity is equal to the number of field lines crossing the light cylinder. We further arbitrarily assume that (i) the number of lines crossing the cylinder can be approximated by the number of pure dipole (non-rotating) lines crossing the same cylinder, and (ii) the lines stay within the same solid angle after crossing the cylinder. Then (\ref{power}) can be approximated by
\begin{equation}
L\propto \int dz d\phi B_{\perp }^2\sin ^2\theta ,
\end{equation}
where the integral is over the light cylinder, and $B_{\perp}$ is the component of the pure-dipole field perpendicular to the light cylinder. Calculating the integral for aligned and orthogonal dipoles, arbitrarily interpolating, and normalizing to the known aligned dipole power, we get $L\approx c^{-3}\mu ^2\Omega^4(1+0.8\sin ^2\chi )$, where $\chi$ is the spin-dipole angle. Obviously, even the sign of the second term is questionable. This estimate, however, may serve to show that the pulsar power is 
\begin{equation}
L\approx c^{-3}\mu ^2\Omega^4,
\end{equation}
for inclinations $\chi \lesssim 30^{\circ}$.

\begin{acknowledgments}
I thank John Bahcall. This work was supported by the David and Lucile Packard Foundation.
\end{acknowledgments}

\begin{appendix}

\section{Isotopological Relaxation}

Consider the problem of minimizing magnetic energy $W$
\begin{equation}
W=\int d^3r B^2,
\end{equation}
by isotopological variations of magnetic field
\begin{equation}
\delta {\bf B}=\nabla \times (\delta \vec{\xi }\times{\bf B}).
\end{equation}
Performing the variation, we get the equation for the minimum energy configuration,
\begin{equation}
{\bf B}\times \nabla \times {\bf B}=0.
\end{equation}
This model is a simplified version of the full pulsar case. 

Since $W$ is positive, any initial configuration can be relaxed into a minimal energy configuration by a series of isotopological deformations. We claim that these deformations turn separatrices into current layers.

\begin{figure}[t]
  \begin{center}
    \includegraphics[angle=0, width=.4\textwidth]{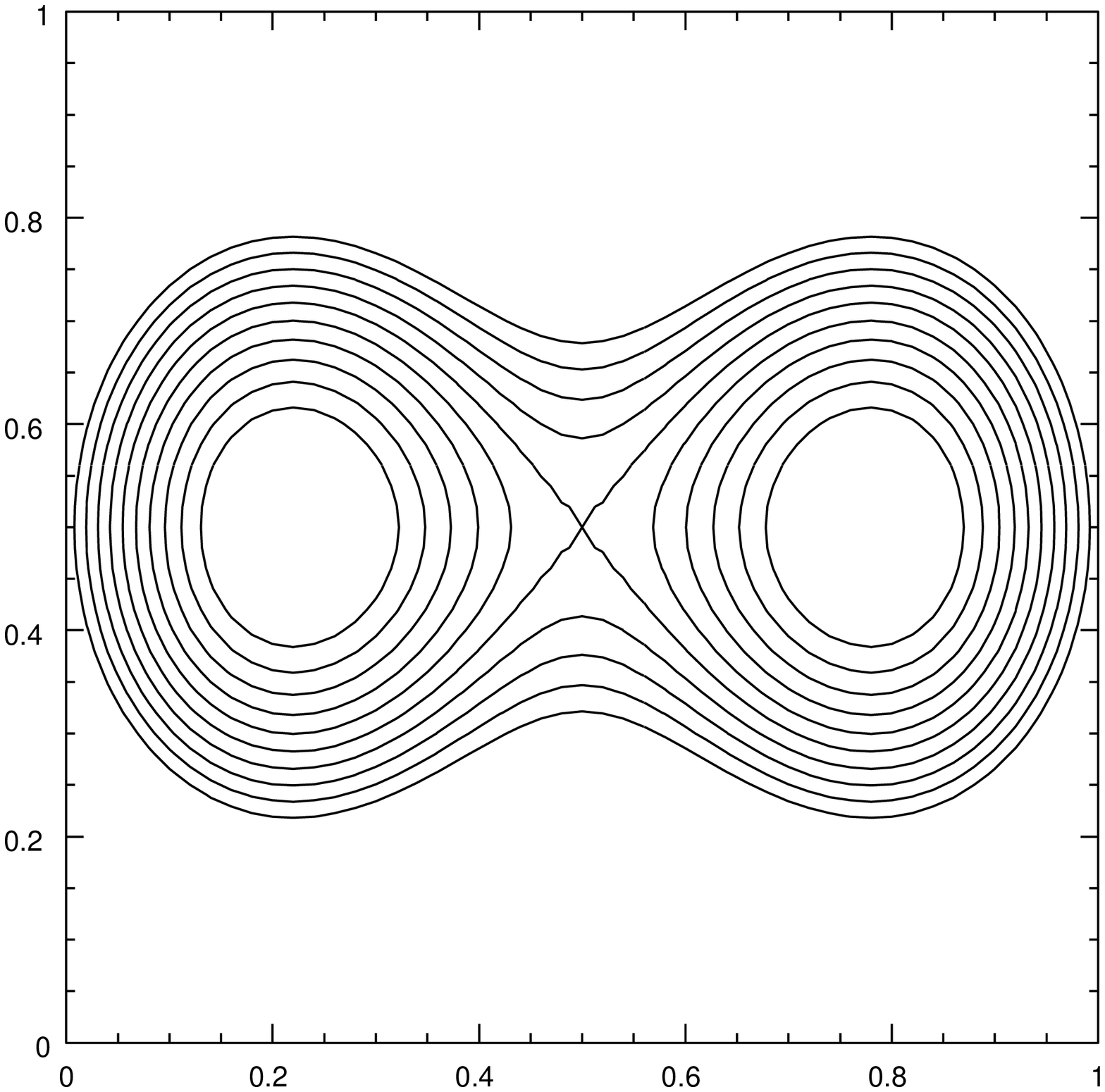}
    \includegraphics[angle=0, width=.4\textwidth]{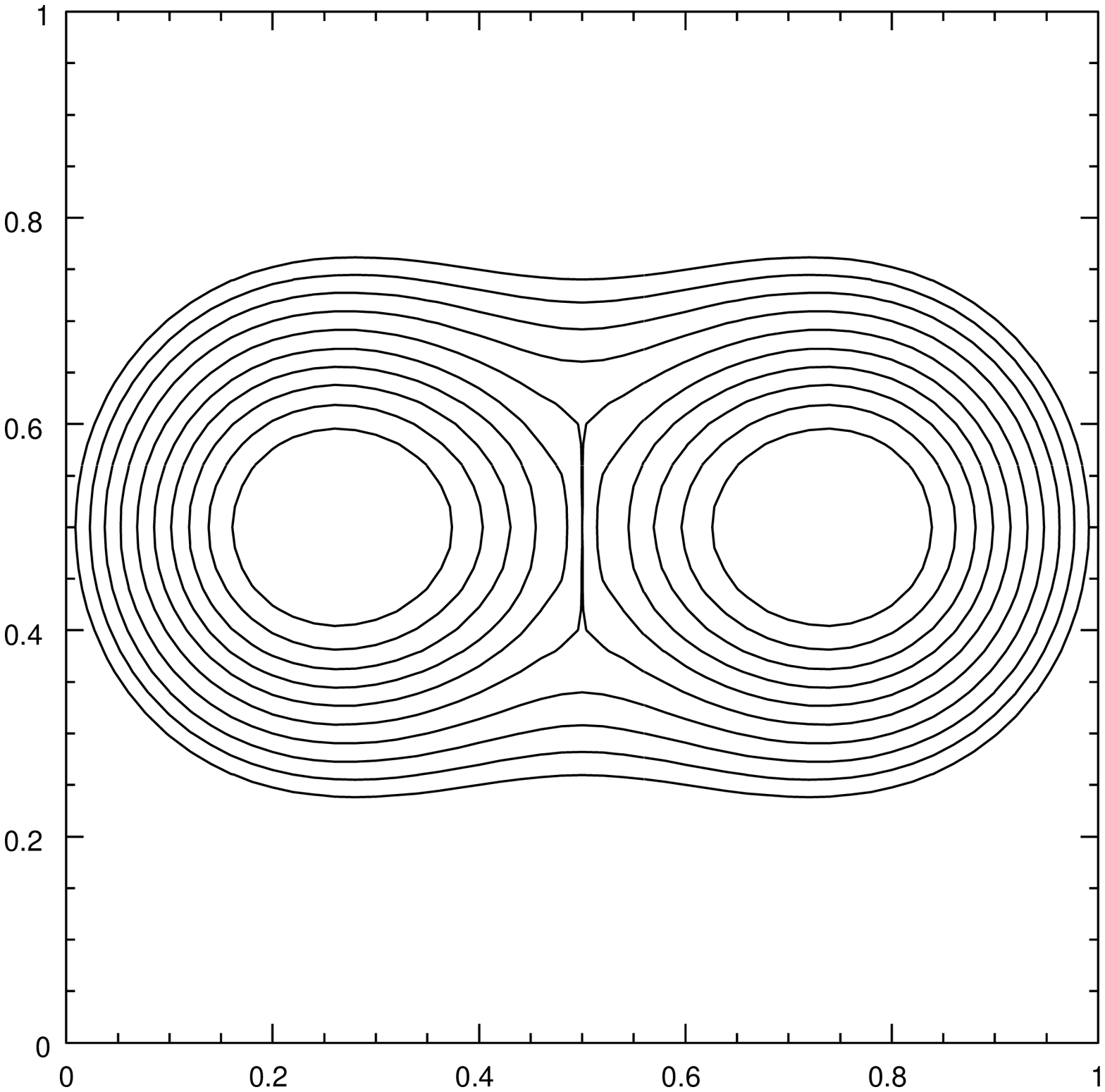}
    \caption{Initial $\psi$ isolines, and $\psi$ isolines some time after the incompressible relaxation (\ref{relax}) was turned on. }
  \end{center}
\end{figure}

A simple rigorous proof can be given after some further simplifications. Consider  two-dimensional magnetic fields ${\bf B}=(-\partial _y\psi , \partial _x\psi, B)$, where $\psi$ and $B$ are functions of $x$ and $y$ but do not depend on $z$. We also assume that minimizing transformations are z-independent. If $B^2\gg (\nabla \psi)^2$, the field first relaxes to a state of nearly uniform $B$ by compressible deformations in the $xy$ plane. Then, the magnetic energy of the plane field 
\begin{equation}
W=\int d^2r (\nabla \psi)^2,
\end{equation}
will be minimized by incompressible $xy$ motions, that is by 
\begin{equation}
\delta \psi =\{ \delta \chi, \psi \} ,
\end{equation}
where $\{ \chi, \psi \} \equiv \partial _x\chi \partial _y\psi -\partial _y\chi \partial _x\psi$ is  2D Jacobian.

One can implement an explicit minimization. For example, 
\begin{equation}\label{relax}
\dot{ \psi }=\{ \chi, \psi \} ,~~~\chi =\{ \psi, \nabla ^2\psi \}
\end{equation}
is an incompressible motion giving $\dot{ W}\leq 0$. We solved (\ref{relax}) numerically. Fig.1 shows how an X-point splits into two Y-points. The inner part of the separatrix is obviously a singular current layer. The outer part is also a singular current layer. To see this, note that $\oint dl |\nabla \psi |^{-1}$ along each $\psi$-isoline is conserved by incompressible motions, while the inner and outer lengths $\oint dl$ of the relaxed separatrix differ.

\begin{figure}[b]
  \begin{center}
    \includegraphics[angle=0, width=.4\textwidth]{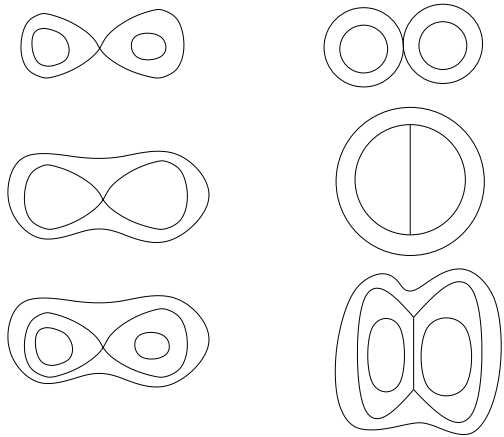}
    \caption{Initial $\psi$ isolines, and $\psi$ isolines after incompressible relaxation. }
  \end{center}
\end{figure}

Fig. 2 gives a simple way to understand the singularity (and, in fact, can be developed into a rigorous proof). If most of energy is in the inner part, the relaxation transforms the separatrix into two circles. If most of energy is in the outer part, the relaxation transforms the separatrix into one circle. Hence, a generic initial state gives a final state with two Y-points.

\end{appendix}

\end{document}